\pdfoutput=1
\documentclass[aps,twocolumn, floatfix, prl]{revtex4}
\usepackage{graphicx}
\DeclareGraphicsExtensions{.pdf}
\usepackage{amsmath,amssymb,bbold,bm,color}
\usepackage{float}
\usepackage{epstopdf}

\newcommand{\bv}{{\bm v}}

\begin{document}

\title{Majorana fermion exchange in strictly one dimensional structures}

\author{Ching-Kai Chiu}
\author{M.M. Vazifeh}
\author{M. Franz}
\affiliation{Department of Physics and Astronomy, University of
British Columbia, Vancouver, BC, Canada V6T 1Z1}
\affiliation{Quantum Matter Institute, University of British Columbia, Vancouver BC, Canada V6T 1Z4}

\begin{abstract}
It is generally thought that adiabatic exchange of  two identical particles is impossible in one spatial dimension. Here we describe a simple protocol that permits adiabatic exchange of two Majorana fermions in a one-dimensional topological superconductor wire. The exchange relies on the concept of ``Majorana shuttle'' whereby a $\pi$ domain wall in the superconducting order  parameter which hosts a pair of ancillary Majoranas delivers one zero mode across the wire while the other one tunnels in the opposite direction. The method requires some tuning of parameters and does not, therefore, enjoy full topological protection. The resulting exchange statistics, however, remain non-Abelian  for a wide range of parameters that characterize the exchange.
\end{abstract}

\date{\today}

\maketitle

Exchange statistics constitute a fundamental property of indistinguishable particles in the quantum theory. In three spatial dimensions general arguments from the homotopy theory constrain the fundamental particles to be either fermions or bosons \cite{leinaas1}, whereas in two dimensions exotic anyon statistics become possible \cite{wilczek1}. In one spatial dimension it is generally believed that statistics are not well defined because it is impossible to exchange two particles without bringing them to the same spatial position in the process. 
Of special interest currently are particles that obey non-Abelian exchange statistics \cite{read2}, both as a deep intellectual challenge and a platform for future applications in topologically protected quantum information processing \cite{nayak1}. Such particles can emerge as excitations in certain interacting many-body systems. In their presence the system exhibits ground state degeneracy and pairwise exchanges of anyons effect unitary transformations on the  ground-state manifold. For non-Abelian anyons the subsequent exchanges in general do not commute and can be used to implement protected quantum computation.

The simplest non-Abelian anyons are realized by Majorana zero modes in
topological superconductors (TSCs)
\cite{read1,ivanov1,kitaev1,alicea_rev,beenakker_rev,stanescu_rev}. In
2D systems they exist in the cores of magnetic vortices while in 1D
they appear at domain walls between TSC and topologically trivial
regions. Although a number of theoretical proposals for 2D
realizations of a TSC have been put forward
\cite{fu1,sau1,alicea1,kallin1} the only credible experimental
evidence for Majorana zero modes thus far exists in 1D semiconductor
wires \cite{sau2,oreg1,mourik1,das1,deng1,finck1,churchill1,Ramon}. Since it is thought
impossible to exchange two Majorana particles in a strictly 1D
geometry, the simplest scheme to perform an exchange involves a three-point turn maneuver in a T-junction comprised of two wires \cite{alicea77} or an equivalent operation \cite{heck1,Flensberg,Sau_exchange} that effectively mimics a 2D exchange. Such
an exchange has been shown theoretically to exhibit the same Ising
statistics that governs Majoranas in 2D systems.  Experimental realization, however,  poses a significant challenge as it requires very high quality T-junctions as well as exquisite local control over the topological state of its segments.  We note that proposals for alternative implementations of Majorana exchange exist that are more realistic \cite{heck1} but still require complex circuitry with multiple quantum wires.
\begin{figure}[b]
\includegraphics[width = 7.6cm]{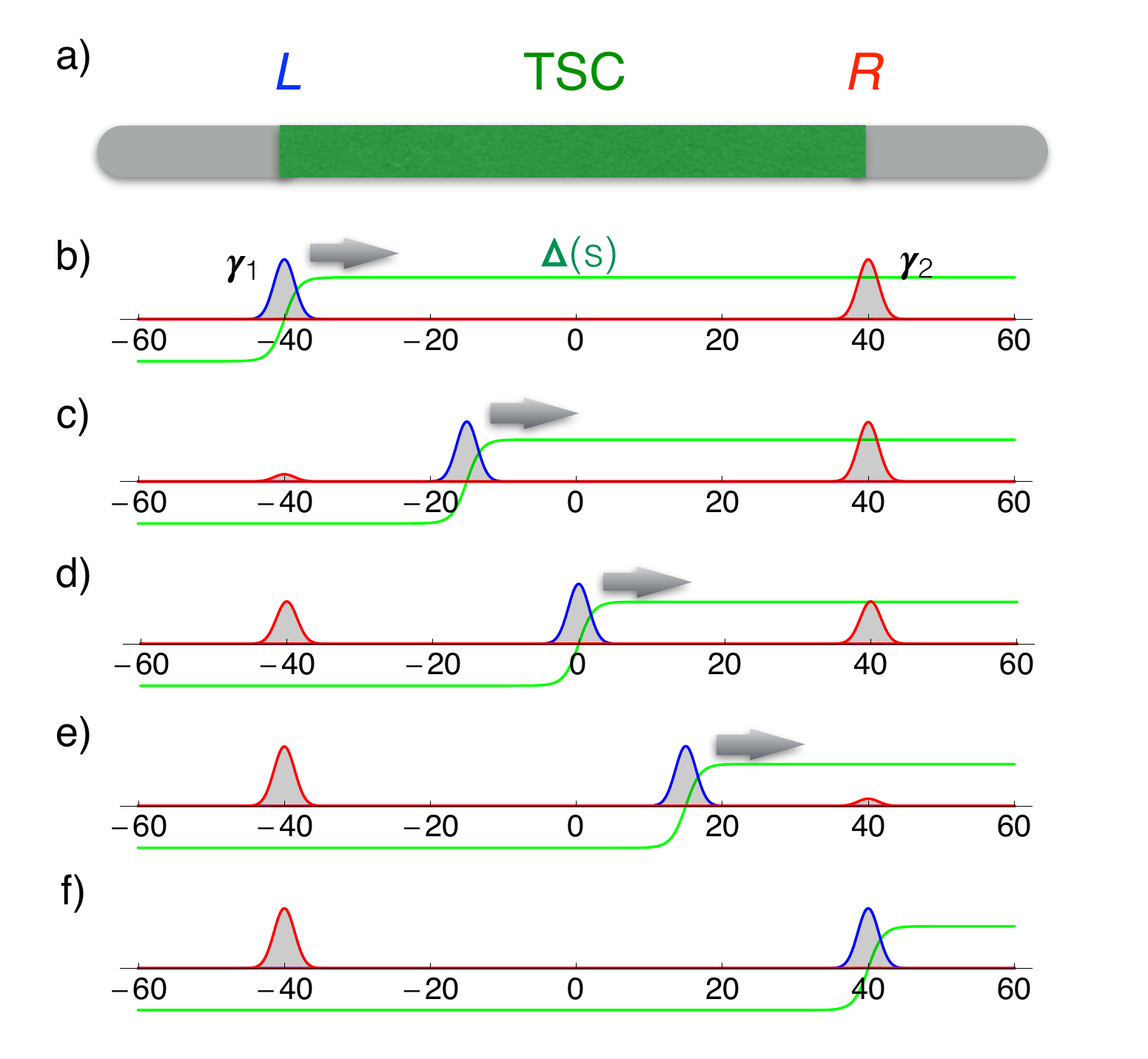}
\caption{Majorana exchange in a 1D wire. a) Schematic depiction of the system with the central green region representing a TSC and outer regions an ordinary superconductor.  Panels b-f) show the Majorana wavefunctions as the kink in $\Delta(s)$ (rendered in green) sweeps the wire from left to right.  
}\label{fig1}
\end{figure}

 In this Letter we introduce a simple protocol that allows for an
 exchange of two Majoranas in a single wire. The basic idea is
 illustrated in Fig.\ \ref{fig1} and relies on the known fact that
 in the presence of an additional symmetry, such as time reversal, a kink in a TSC wire (defined as a $\pi$ domain wall in the phase of the superconducting order parameter) carries a protected {\em pair} of Majorana zero modes \cite{note2,fu79,tewari2,simon1,ojanen1}. As we show in detail below, under a wide range of conditions such a kink acts as a transport vehicle (``shuttle'') for the Majorana end modes. Specifically, as the kink traverses the wire from left to right it brings along with it the left end-mode (Fig.\ \ref{fig1}b,c). Meanwhile, the right end mode tunnels through the wire to its left end    (Fig.\ \ref{fig1}d-e) so that the end result is an adiabatic exchange of two Majorana zero modes (Fig.\ \ref{fig1}f). We show that this exchange satisfies the usual rules of Ising braid group, namely
\begin{equation}\label{ex1}
\gamma_1\mapsto\gamma_2, \ \ \ \ \gamma_2\mapsto-\gamma_1.
\end{equation}
We also discuss various limitations of our exchange protocol and its possible physical realizations.

Consider a process in which a  $\pi$ domain wall is nucleated in the quantum wire Fig.\ \ref{fig1}a to the left of point $L$ and is then transported to the right along the wire. Initially, the  domain wall is in the trivial superconductor and we do not expect it to bind any zero modes. We model this situation by the low-energy Hamiltonian
\begin{equation}\label{h0}
H_0=i\epsilon\Gamma_1\Gamma_2,
\end{equation}
where $\Gamma_j$ denote the Majorana operators that we anticipate to become zero modes once the kink reaches the TSC while $\epsilon$ is their energy splitting in the trivial phase.  As the kink approaches the topological segment of the wire we expect two things to happen: (i) $\Gamma_j$ become zero modes, meaning we should set $\epsilon\to 0$ and (ii) since their wavefunctions begin to overlap with the wavefunction of $\gamma_1$, we expect terms $i\gamma_1\Gamma_j$ to enter (\ref{h0}) with non-zero coefficients. 

Since the Hamiltonian (\ref{h0}) is invariant under a rotation in $(\Gamma_1,\Gamma_2)$ space, i.e.
$\Gamma_1\to\Gamma_1\cos{\alpha}+\Gamma_2 \sin{\alpha}$ and $\Gamma_2\to\Gamma_2\cos{\alpha}-\Gamma_1 \sin{\alpha}$ we may, without loss of generality, select  $\alpha$ such that the resulting Hamiltonian reads
\begin{equation}\label{h1}
H_{\rm I}=i\epsilon\Gamma_1\Gamma_2 +it_1\gamma_1\Gamma_2.
\end{equation}
This Hamiltonian reveals an important property of the system that will be key to the functioning of our device: as $\epsilon$ decays to zero and $t_1$ ramps up to a non-zero value, the zero mode denoted as $\gamma_1$ originally positioned at $L$ transforms into zero mode $\Gamma_1$ located at the kink. Physically, the kink subsumes the Majorana zero mode and {\em transports it along the wire} as illustrated in \ref{fig1}b,c. This is the key idea behind the Majorana shuttle.

As the kink approaches the center of TSC we can no longer ignore coupling $t_2$ to the other end mode denoted by $\gamma_2$. Indeed when the kink is exactly midway we expect $t_1=t_2$ on the basis of symmetry. The relevant Hamiltonian then becomes 
\begin{equation}\label{h2}
H_{\rm II}=it_1\gamma_1\Gamma_2+it_2\gamma_2\Gamma_2.
\end{equation}
As the kink advances along the wire and $t_1$ declines while $t_2$ grows the zero mode $\gamma_2$ transforms into $\gamma_1$. Physically, the zero mode initially located at $R$ tunnels across the length of the TSC and reappears on the other side at $L$. 

Finally, as the kink traverses into the trivial phase to the right of the TSC the pair of ancillary Majoranas acquire a gap and $t_2$ reaches zero. We describe this by  
\begin{equation}\label{h3}
H_{\rm III}=i\epsilon\Gamma_1\Gamma_2 +it_2\gamma_2\Gamma_2.
\end{equation}
As before, this shows that $\Gamma_1$ transforms into $\gamma_2$, completing the exchange.

\begin{figure}[b]
\includegraphics[width = 7.6cm]{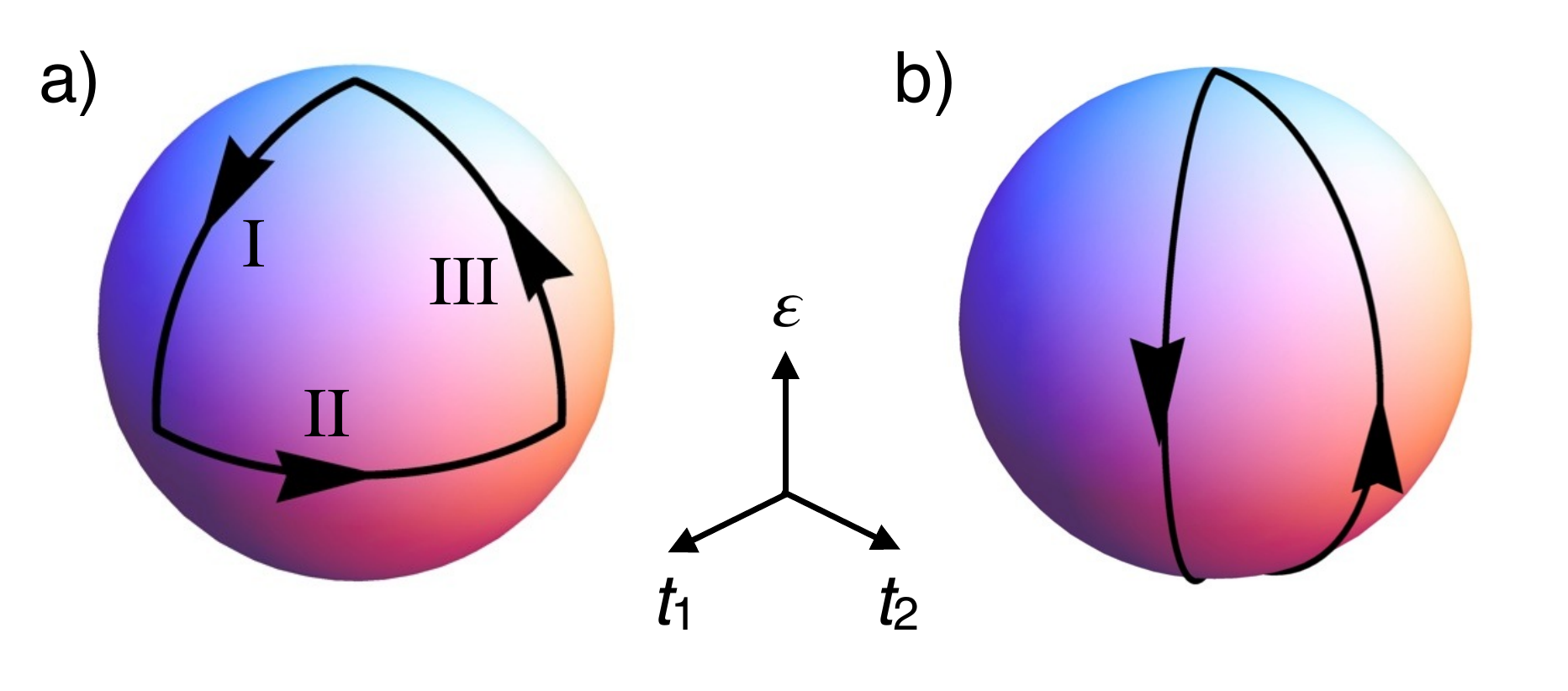}
\caption{Parameter path ${\bf h}(s)$ on the unit sphere. a) The path representing Majorana shuttle exchange Eq.\ (\ref{sph3}). b) An alternate exchange protocol discussed in Supplementary Material.  
}\label{fig2}
\end{figure}
The entire sequence of events can be described economically by a single Hamiltonian \cite{note1} 
\begin{equation}\label{hs}
H(s)=i[t_1(s)\gamma_1 +t_2(s)\gamma_2+\epsilon(s)\Gamma_1]\Gamma_2.
\end{equation}
where $s$ represents the kink position along the wire, increasing from left to right. 
The spectrum of $H(s)$ consists of two exact
zero modes and two non-zero energy levels $\pm
E(s)=\pm\sqrt{t_1^2+t_2^2+\epsilon^2}$. Since we expect Majorana
wavefunctions to decay exponentially at long distances a reasonable
assumption inside TSC is $\epsilon(s)=0$ and $t_{1,2}(s)\approx t_0 e^{(\pm s-l/2)/\xi}$
where $l$ is the TSC length, $\xi$ represents the decay length of the
Majorana wavefunctions and $s$ is referenced to the TSC midpoint. This gives
\begin{equation}\label{es}
E(s)\approx 2t_0 e^{-l/2\xi}\sqrt{\cosh{(2s/\xi)}}.
\end{equation}

It is useful to write the three couplings as a vector ${\bf h}(s)=(t_1(s),t_2(s),\epsilon(s))$, in spherical coordinates
\begin{equation}\label{sph1}
{\bf h}(s)
=E(s)(\sin{\theta}\cos{\varphi},\sin{\theta}\sin{\varphi},\cos{\theta})
\end{equation}
with angles $\theta$ and $\varphi$ dependent on $s$. The exchange can
thus be visualized as a path on the unit sphere Fig.\ \ref{fig2}a 
parametrized by
$[\theta(s),\varphi(s)]$; the amplitude $E(s)$ is unimportant as long
as it remains non-zero. 

\begin{figure*}[t]
\includegraphics[width = 17.6cm]{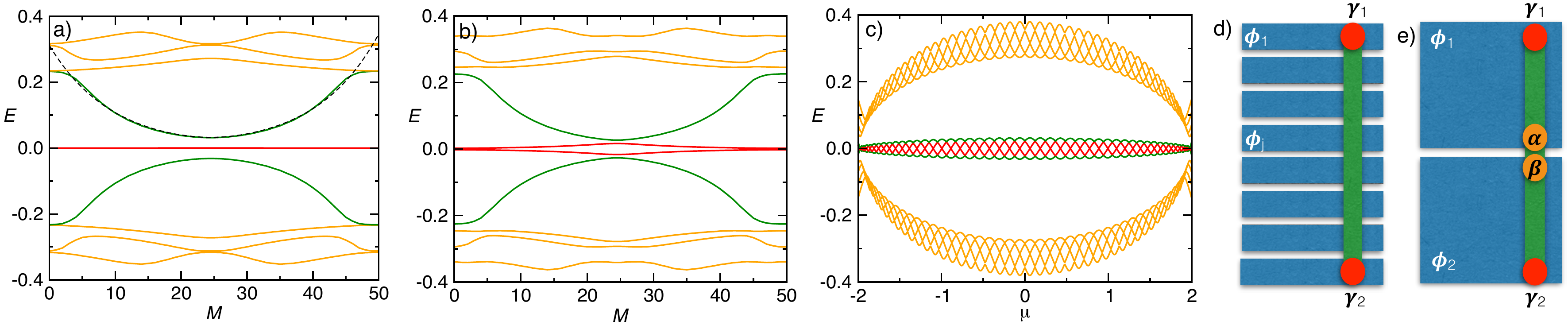}
\caption{Lowest lying energy levels of the lattice Hamiltonian (\ref{hkit1}) with $\Delta=0.1t$ and $N=50$. a) Shown as a function of the kink position $M$ for $\mu=0.185t$ and b) for $\mu=0.140t$. The dashed line in panel a) represents  the best fit of the second energy level to Eq.\ (\ref{es}) with $t_0=0.246t$ and $\xi=9.20$.
Panel c) displays the same energy levels as a function of the chemical potential $\mu$ with the kink fixed at the TSC midpoint $M=25$. Panels d,e) illustrate possible physical realizations of our device. Blue areas indicate SC electrodes with controlled phases $\phi_j$, green is the quantum wire with Majorana endmodes indicated in red.
}\label{fig3}
\end{figure*}
The zero mode operators $\gamma(s)$ of the Hamiltonian (\ref{hs}) satisfy 
$[H(s),\gamma(s)]=0$ and correspond to two vectors $\bv_1(s)$ and $\bv_2(s)$ orthogonal to ${\bf h}(s)$. They read
\begin{eqnarray}\label{sph2}
\gamma_1(s)&=& \gamma_1\cos{\theta}\cos{\varphi}+\gamma_2\cos{\theta}\sin{\varphi}  - \Gamma_1 \sin{\theta}\nonumber \\
\gamma_2(s)&=&-\gamma_1\sin{\varphi}+\gamma_2\cos{\varphi}
\end{eqnarray}
Of course any linear combination of $\gamma_1(s)$ and $\gamma_2(s)$ is also a zero mode of $H(s)$. The zero modes that solve the appropriate time-dependent Schr\"{o}dinger equation are those linear combinations which make the non-Abelian Berry matrix  $A_{ab}=\bv_a\cdot\partial_s\bv_b$ vanish \cite{zee1}. With this in mind we can track the evolution of the zero modes as the couplings change. We do this in three stages as described by Hamiltonians $H_{\rm I}$, $H_{\rm II}$ and $H_{\rm III}$ above. The spherical angles evolve as
\begin{eqnarray}\label{sph3}
\theta: &&\ \  0\stackrel{\rm I}\longrightarrow {\pi\over 2} \stackrel{\rm II}\longrightarrow {\pi\over 2} \stackrel{\rm III}\longrightarrow 0,
\nonumber \\
\varphi: && \ \  0\stackrel{\rm I}\longrightarrow 0 \stackrel{\rm II}\longrightarrow {\pi\over 2} \stackrel{\rm III}\longrightarrow {\pi\over 2},
\end{eqnarray}
which implies, according to Eq.\ (\ref{sph2}), the following evolution of the Majorana zero modes
\begin{eqnarray}\label{ex2}
\gamma_1(s): && \ \ \gamma_1\stackrel{\rm I}\longrightarrow -\Gamma_1 \stackrel{\rm II}\longrightarrow -\Gamma_1 \stackrel{\rm III}\longrightarrow \gamma_2,
\nonumber \\
\gamma_2(s): && \ \  \gamma_2\stackrel{\rm I}\longrightarrow \ \ \gamma_2 \stackrel{\rm II}\longrightarrow -\gamma_1 \stackrel{\rm III}\longrightarrow -\gamma_1.
\end{eqnarray}
We observe that the exchange protocol indeed implements the Ising braid group Eq.\ (\ref{ex1}).

The result shown in Eq.\ (\ref{ex2}) is a direct consequence of the structure  of the Hamiltonian (\ref{hs}) and is in that sense exact. However,
in the derivation leading to our main result (\ref{ex2}) we made an important assumption that in Hamiltonian (\ref{hs}) $\gamma_2$ only couples to $\Gamma_2$. This is non-generic because  a term $it_2'\gamma_2\Gamma_1$ is also allowed and cannot be removed by a rotation in $(\Gamma_1,\Gamma_2)$ space without generating additional undesirable terms. Such a coupling, when significant, spoils the exact braiding property Eq.\ (\ref{ex2}) because it shifts the Majorana zero modes to non-zero energies $\pm t_2'$ during step II. Parameter $t_2'$ (like the remaining parameters in $H(s$)) depends on Majorana wavefunction overlaps and is therefore non-universal. In the following we study a simple lattice model for a TSC and show that the situation with   $|t_2'|$ much smaller than all the other relevant parameters can be achieved by tuning a single system parameter such as the chemical potential $\mu$ or the length $l$ of the topological segment of the wire. The necessity to tune $t_2'$  to zero is the price one must pay in order to exchange Majoranas reliably in a 1D wire.

To put our ideas to the test we now study the $\pi$ kink in the prototype lattice model of TSC due to Kitaev \cite{kitaev1}. It describes spinless fermions hopping between the sites of a 1D lattice defined by the  Hamiltonian 
\begin{equation}\label{hkit1}
{\cal H_{\rm latt}}=\sum_j\left[(-t c^\dagger_j c_{j+1} +\Delta_{j,j+1} c^\dagger_j c^\dagger_{j+1}+{\rm h.c.}) - \mu  c^\dagger_j c_{j}\right]
\end{equation}
where $\Delta_{j,j+1}$ is the superconducting order parameter on the bond connecting sites $j$ and $j+1$. For non-zero uniform $\Delta_{j,j+1}$ the chain is known to be in a TSC phase when $|\mu|<|2t|$. Here we study  an open ended chain with $N$ sites and a kink described as 
\begin{equation}\label{kink}
\Delta_{j,j+1}=\biggl\{
\begin{matrix}
-\Delta, \ \ j\leq M \\
+\Delta, \ \ j> M
\end{matrix}
\end{equation}
In the limit of a long wire it is possible to find various zero mode wavefunctions analytically and from their matrix elements derive the effective Hamiltonian Eq.\ (\ref{hs}). Since the details of such calculations are tedious and not particularly illuminating we focus here on exact numerical diagonalizations which convey the key information with greater clarity.
Figure \ref{fig3}a,b shows the energy eigenvalues of ${\cal H_{\rm latt}}$
as a function of the kink position $M$ for two typical and physically distinct situations. The four eigenvalues closest to zero (rendered in green and red) are associated  with the Majorana modes
$\gamma_1, \gamma_2, \Gamma_1$ and $\Gamma_2$ discussed above. In Fig. \ref{fig3}a
we observe two near-zero modes (red) and two modes at finite energy behaving as
expected from Eq.\ (\ref{es}). For these chosen
parameters the system behaves in accordance with our effective low-energy theory defined by the Hamiltonian
$H(s)$ in Eq.\ (\ref{hs}). The inspection of the associated
wavefunctions indeed confirms behavior indicated in Fig.\ \ref{fig1}b-f,  fully consistent with the Majorana shuttle concept.  For a slightly
shifted chemical potential $\mu$ Fig. \ref{fig3}b shows that the zero
modes are lifted, indicating that coupling $t_2'$ has become
significant. In this case the low-energy theory (\ref{hs}) does not 
apply and the adiabatic exchange is compromised.  

To better understand the interplay between
the two types of behavior contrasted in Fig.\ \ref{fig3}a,b  we plot in panel c the energy spectrum of ${\cal H_{\rm latt}}$ as a function of
the chemical potential for the kink fixed at the TSC midpoint $(M=N/2)$ where the difference is most pronounced. 
The oscillatory behavior of the energy levels here reflects the fact that in
addition to the simple exponential decay Majorana wavefunctions also
exhibit oscillations at the Fermi momentum $k_F$ of the underlying normal
metal. These oscillations affect the wavefunction overlaps and
thus influence the coupling constants in the effective Hamiltonian.
If we denote the two lowest non-negative eigenvalues  of ${\cal H_{\rm latt}}$ by $E_0(\mu)$ and $E_1(\mu)$ then, for the Majorana shuttle to function, we require that
\begin{equation}\label{emu}
E_0(\mu)\ll E_1(\mu).
\end{equation}
When (\ref{emu}) is satisfied then one can perform the exchange operation
sufficiently fast compared to $\hbar/E_0(\mu)$  so that the small energy splitting between the zero modes
does not appreciably affect the result and at the same sufficiently
slow compared to $\hbar/E_1(\mu)$ so that the condition of adiabaticity is satisfied and the system
remains in the ground state. If on the other hand $E_0(\mu)$ and
$E_1(\mu)$ are comparable then such an operation becomes impossible.
In the wires used in the Delft experiment \cite{mourik1} 
the size of the Majorana wavefunction $\xi$ was estimated to be about $1/10$ of the
wire length $l$. When the kink is near the wire midpoint one may thus
crudely estimate  $E_1(\mu)\simeq\ t(l/2\xi) e^{-l/2\xi}\simeq
0.3\Delta$. Parameters in our numerical simulations were chosen to
yield comparable values. One may similarly estimate the typical
maximum value of $E_0(\mu)$ for the situation when $\mu$ is
appropriately tuned. This value would be zero in the ideal clean case
and if we ignore the mutual overlap of the end-mode Majorana
wavefunctions. Considering the latter to be non-zero we obtain $E_0(\mu)\simeq\ t(l/\xi) e^{-l/\xi}\simeq
0.0045 \Delta$. We may thus conclude that with the existing wires and under favorable conditions the
adiabaticity requirement $E_0(\mu) \ll E_1(\mu)$ can be satisfied with
at least an order of magnitude to spare.

Inspection of Fig.\ \ref{fig3}c further reveals that a small adjustment of the
average chemical potential $\mu$ should be sufficient in most cases to
tune the system to satisfy Eq.\ (\ref{emu}). In essence, we require $\mu$ to be tuned close to one of the zeroes $\mu_n$ of $E_0(\mu)$. From Fig.\ \ref{fig3}c one can deduce a simple heuristic formula
\begin{equation}\label{emu0} 
E_0(\mu)\simeq f(\mu)\min(|\sin{k_F l}|,|\cos{k_F l}|)
\end{equation}
with $f(\mu)$ a slowly varying envelope function. The zeroes occur at $k_Fl={\pi\over 2}n$ with $n$ integer from which one can infer the spacing between the successive values of $\mu_n$ as  $\delta\mu\approx\pi\mu/2 N$. 
We see that if $\mu$ is initially set at random, for a long wire only a minute adjustment (achieved e.g.\ by gating) is required to bring it to the regime where adiabatic exchange using the Majorana shuttle protocol becomes feasible.
In the Supplementary Material we furthermore show that the exchange protocol is robust with respect to moderate amounts of disorder in the chemical potential $\mu$, hopping $t$ and pairing amplitude $\Delta$. Specifically, the exchange occurs as long as the fluctuations in the above parameters do not significanly exceed the average pairing amplitude $\Delta_0$.   

The Majorana shuttle can be physically realized by coupling a single semiconductor wire \cite{sau2,oreg1,mourik1,das1,deng1,finck1,churchill1} to a ``keyboard'' of superconducting electrodes as illustrated in Fig.\ \ref{fig3}d. If one can control the phases $\phi_j$ of the individual electrodes (e.g.\ by coupling them to flux loops) then a phase kink can be propagated along the wire in steps, implementing the proposed exchange protocol. It might be also possible to generate the kink by running a current between the adjacent electrodes in a variant of the setup discussed in \cite{romito1}; large enough supercurrent should produce a phase difference close to $\pi$ owning to the fundamental Josephson relation $I(\Delta\phi)=I_0\sin{\Delta\phi}$.

An even simpler physical realization follows from the generalized exchange protocol discussed in the Supplementary Material. There we demonstrate that, as a matter of fact, an exchange of $\gamma_1$ and $\gamma_2$ is effected by {\em any} closed parameter path ${\bf h}(s)$ in the Hamiltonian (\ref{hs}) provided that (i) it starts at the pole of the unit sphere Fig.\ \ref{fig2} and (ii) it sweeps a solid angle ${\pi\over 2}$. One such path, shown in Fig.\ \ref{fig2}b, can be physically realized in a setup with only two SC electrodes \citep{San-Jose} indicated in Fig.\ \ref{fig3}e by simply twisting the phase of one of the electrodes by $2\pi$ (see Supplementary Material for details). As before, we require that endmodes $\gamma_1$ and $\gamma_2$ couple to the same ancillary Majorana, say $\Gamma_2$, localized in the juction. In addition, for the exchange to obey Eq.\ (\ref{ex1}), it is necessary that $t_1=t_2$ to a good approximation, which can be implemented by positioning the junction midway along the wire.  

Finally, we note that a pair of ancillary Majoranas is also realized at a magnetic domain wall in the chain of magnetic atoms deposited on the surface of a superconductor \cite{nadj-perge2,ojanen1}. Our exchange protocol works equally well in this situation, if a way can be found to manipulate the domain wall.

Our considerations demonstrate that it is in principle possible to exchange two Majorana fermions in a strictly one-dimensional system. The price one pays for this is the necessity to tune a single parameter (e.g.\ the global chemical potential) in the device. In the Majorana shuttle protocol one must in addition impose a symmetry constrain (such as the time reversal) to protect the Majorana doublet at the domain wall but this condition is relaxed in the Josephson junction implementation.  Our proposed protocol involves four Majoranas, two of them ancillary, and relies on the fact that a pair of exact zero modes is preserved when only three of the four are mutually coupled. In steps I and III of the exchange this is guaranteed by virtue of the fourth Majorana being spatially separated from the remaining three. In step II one must tune a parameter to achieve the decoupling of the fourth Majorana. In this respect our protocol is similar to Coulomb assisted braiding \cite{heck1} but is potentially simpler to implement because it involves only a single quantum wire.      

The authors are indebted to  NSERC, CIfAR and Max Planck - UBC Centre for Quantum Materials for support.
%...


\begin{thebibliography}{10}
\bibitem{note2} A 1D system obeying both the particle-hole and time-reversal symmetry is in BDI topological class characterized by an integer-valued invariant $\nu$. Segments with the opposite sign of the SC order parameter have $\nu=\pm 1$ and a domain wall thus binds two Majorana fermions.
\bibitem{note1} We note that this Hamiltonian has been studied previously in the context of Coulomb assisted braiding in Ref.\ \cite{heck1}.
\end{thebibliography}
\end{document}

\title{Supplementary Material for ``Majorana fermion exchange in strictly one dimensional structures''}
\author{Ching-Kai Chiu}
\author{M.M. Vazifeh}
\author{M. Franz}
\affiliation{Department of Physics and Astronomy, University of
British Columbia, Vancouver, BC, Canada V6T 1Z1}
\affiliation{Quantum Matter Institute, University of British Columbia, Vancouver BC, Canada V6T 1Z4}

\date{\today}
\maketitle

\appendix
\section{Effect of disorder and Berry matrix evaluation}

\setcounter{figure}{3}
\begin{figure*}
\includegraphics[height = 5cm]{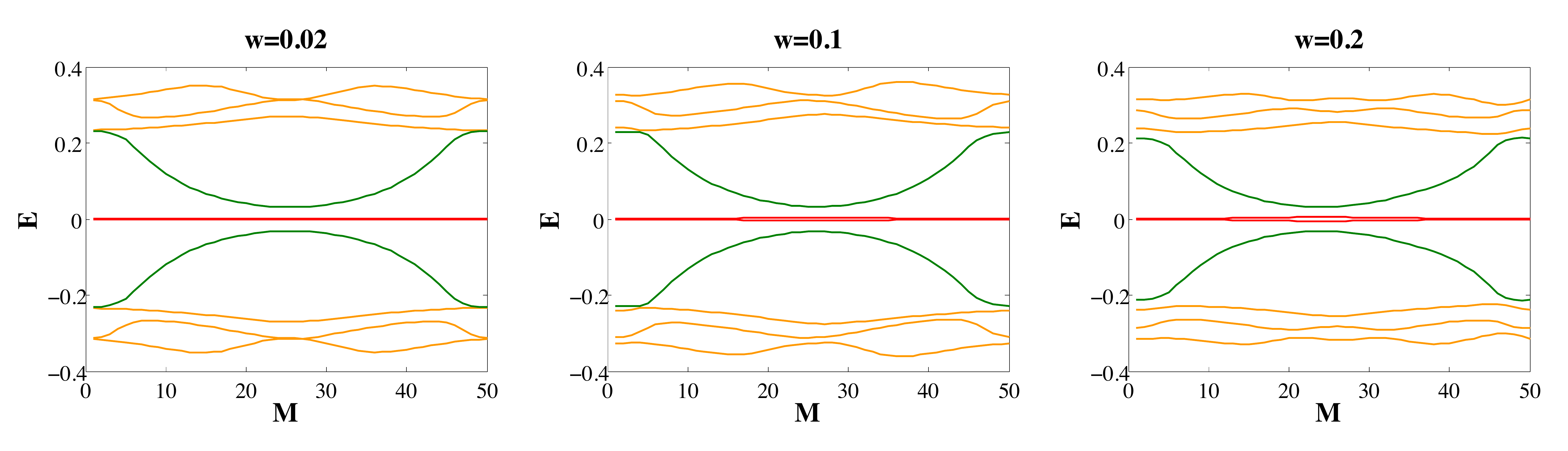}
\caption{The effect of on-site disorder on the spectrum of the Kitaev chain with a $\pi$ domain wall. Parameters are as in Fig.\ 3a of the manuscript, namely $\Delta=0.1t$, $\mu=0.185t$  and $N=50$. The stronger disorder, the smaller energy gap between the ground states and the excited states in average. }  
\label{fig2}
\end{figure*}
In order to address the stability of our proposed exchange protocol we
have performed additional numerical simulations of the Kitaev chain
with a $\pi$ domain wall in the presence of  disorder in the on-site potential and other parameters. To this end we add a disorder term 
\begin{equation}\label{hdis}
{\cal H_{\rm dis}}=\sum_j \delta\mu_j  c^\dagger_j c_{j}
\end{equation}
to the lattice Hamiltonian ${\cal H_{\rm latt}}$ defined in Eq.\ (12)
of the manuscript. Here $\delta\mu_j$ is a random potential uniformly
distributed in the interval $(-w,w)$. The characteristic spectra of
the Hamiltonian as a function of the domain wall position $M$ are
shown in Fig.\ \ref{fig2} for various values of disorder strength
$w$. We observe that weak disorder $w=\Delta/5=0.02t$ has no visible
effect on the zero modes. Even relatively strong disorder
$w=\Delta=0.1t$ has only modest effect on the zero modes. Only in the
dirty limit, i.e.\ when  $w$ significantly exceeds $\Delta$, we find a notable splitting between the
zero mode energies. 
We have also investigated the effect of disorder in the hopping
amplitudes $t$ and the pairing potential amplitude $\Delta$ with similar
results.

Of particular interest is our result in Fig.\ \ref{fig3}
showing that {\em phase} disorder in
$\Delta_{j,j+1}=\Delta_0e^{i\delta\phi_{j,j+1}}$  has no visible
effect on the zero modes, despite the fact that complex order parameter breaks
the time reversal symmetry of the problem that protects the doublet of
the ancillary majoranas. From these simulations it appears that the time
reversal symmetry must only be preserved on average to protect the
Majorana zero modes residing at the domain wall. Although we do not
fully understand the fundamental basis for this result we have
verified it to be true numerically for many different realizations of
the phase disorder.

Results of our simulations with various types of disorder indicate that the exchange protocol proposed in our manuscript is robust. Given the system with a specific realization of disorder one can tune a single parameter, such as the global chemical potential $\mu$, until the splitting between the zero modes is minimized, as indicated in Fig.\ \ref{fig3}. This is possible so long as the disorder strength does not significantly exceed the size of the superconducting gap $\Delta$. Once again, such a limitation on the disorder strength is common to all 1D realizations of Majorana zero modes: too strong a disorder would ultimately destroy the topological phase.

In addition to the quasiparticle spectra, which we showed to support our low-energy effective theory, it is possible to obtain a more detailed characterization of the Majorana exchange from the numerical simulation of the Kitaev chain. Below, we directly evaluate the non-Abelian Berry matrix introduced by Wilczek and Zee \cite{zee1} which describes the unitary evolution of the zero modes in the degenerate ground state manifold. According to Ref.\ \onlinecite{zee1}, the solution of the time-dependent Schr\"odinger equation 
\begin{equation}\label{schr}
i{\partial\psi\over\partial t}=H(t)\psi
\end{equation}
for degenerate states $\psi_a$ in the adiabatic limit is given by 
\begin{equation}\label{unit1}
\psi_a(t)=U_{ab}(t)\psi_b(0).
\end{equation}
Here 
\begin{equation}\label{unit2}
U_{ab}(t)=\exp{\left[\int_0^t\langle\eta_a(\tau)|\dot{\eta}_b(\tau)\rangle d\tau +
\ln{\langle\eta_a(t)|{\eta}_b(0)\rangle}\right]}
\end{equation}
represents the unitary evolution operator (the ``Berry matrix'') and $\eta_b(t)$ are instantaneous degenerate eigenstates of $H(t)$ whose dependence on $t$ is chosen to be smooth and to satisfy the initial condition $\eta_a(0)=\psi_a(0)$.

Our numerical diagonalization of the Kitaev chain Hamiltonian yields the instantaneous eigenstates $\eta_{1,2}(t)$ associated with the Majorana zero modes for the domain wall position $M$ at time $t$. We can use these to evaluate $U_{ab}(t)$ from Eq.\ (\ref{unit2}) and thus ascertain the unitary evolution of the Majorana zero modes described by $\psi_a(t)$. In practice, we always encounter a small energy splitting between the zero modes due to the finite size of our system and potentially other effects such as the small undesirable coupling $t'_2$. In the presence of a non-zero energy splitting we choose the instantaneous eigenstates $\eta_{1,2}(t)$ as linear combinations of
\begin{figure}[b]
\includegraphics[width = 8.6cm]{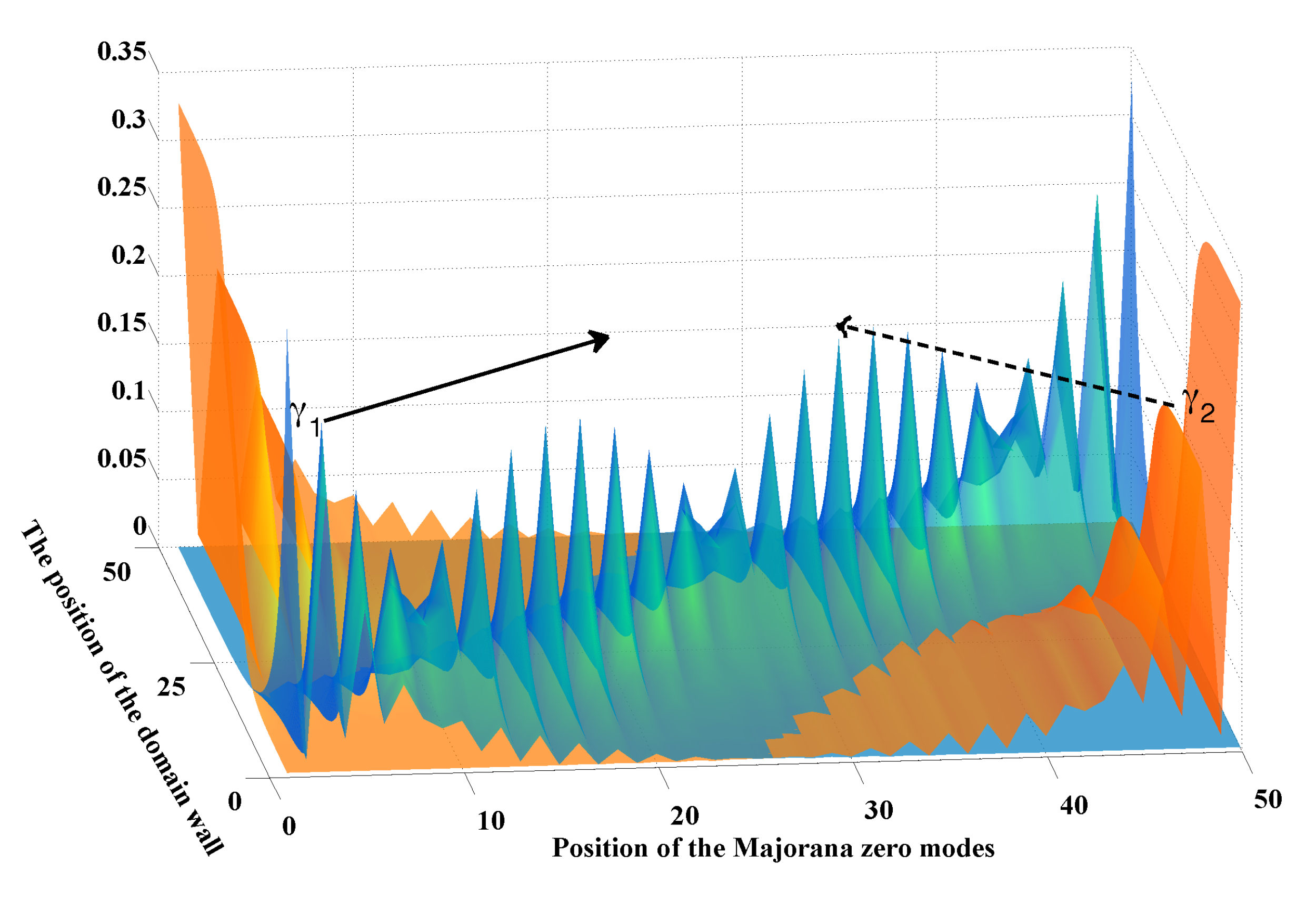}
\caption{Majorana wavefunction amplitudes $|\eta_{1,2}(t)|^2$ as chosen for the Berry matrix evaluation in the Kitaev chain with a $\pi$ domain wall. As the domain wall moves from $0$ to $50$, the Majorana mode $\gamma_1$ continuously moves from $0$ to $50$. When the domain wall passes through site $25$, the other Majorana mode ($\gamma_2$) \emph{teleports} from the left end to the right end.}
\label{wf}
\end{figure}
 the two low-energy states that (i) obey the desired initial condition (i.e.\ represent two Majorana particles at the ends of the wire), and (ii) evolve smoothly as we move the domain wall while their wavefunctions remain Majorana throughout the process. An example of the wavefunctions chosen in this way is given in Fig.\ \ref{wf}. With this choice of basis we typically find that the integral in the exponent of Eq.\ (\ref{unit2}) vanishes while the second (monodromy) term is nonzero. Together this gives
\begin{equation}\label{unit3}
U_{ab}(t)\simeq
\begin{pmatrix}
0 & 1 \\
-1 & 0
\end{pmatrix},
\end{equation}
to within the numerical accuracy of our calculation. Once again this confirms that the two Majorana zero modes exchange in the process and that the exchange satisfies the rules of the Ising braid group given in Eq.\ (1) of the manuscript. Importantly, the result in Eq.\ (\ref{unit3}) remains valid in the presence of moderate amounts of disorder.
\begin{figure*}
\includegraphics[height = 5cm]{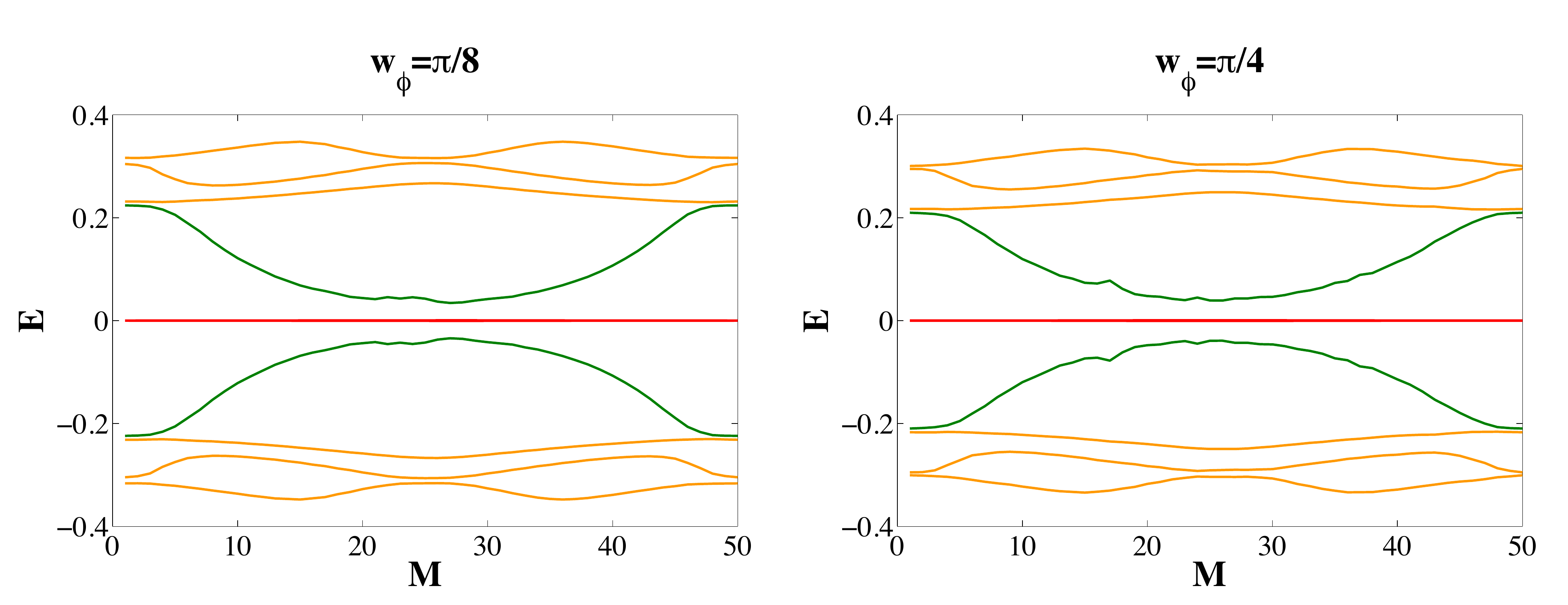}
\caption{The effect of phase disorder on the spectrum of the Kitaev chain with a $\pi$ domain wall. Random phase $\delta\phi_{j,j+1}$ is uniformly distributed between $(-w_\phi,w_\phi)$.}
\label{fig3}
\end{figure*}
%

%%%%%%%%%%%%%%%%%%%%%%%%%%%%%%%%%%%%%%%%%%%%%%%%%%%%%%%%%%%%%%%%%%

\section{Generalized exchange protocol}\label{generalization}
In strictly one dimension the Majorana coupling path indicated in Fig.\ 2a in the $\pi$ domain wall protocol implements an exchange between $\gamma_1$ and $\gamma_2$. This path is not the only one that achieves braiding. In this section, we discuss other paths in the parameter space of the Hamiltonian (6) that can bring about braiding.  Before the general discussion, we consider another specific coupling path of the same four Majorana fermions ($\gamma_1,\ \gamma_2,\ \Gamma_1,\Gamma_2$) designed to achieve braiding of $\gamma_1$ and $\gamma_2$. As we shall argue below this path can be implemented by the setup displayed in Fig.\ 3e.

The path we consider consists of two steps and is depicted in Fig.\ 2b. We start from $H=i\epsilon \Gamma_1\Gamma_2$ ($\theta=0,\ \varphi=0$) and let the coupling vary from $\theta=0$ to $\theta=\pi$ as $\varphi=0$ remains fixed. As shown in Fig.\ 2b, the point on the unit sphere moves straight from the north pole to the south pole. According to Eq.\ (9) this produces the following evolution of the zero modes
\begin{align}
\gamma_1 \rightarrow -\gamma_1,\quad \gamma_2 \rightarrow \gamma_2. 
\end{align}
Now the coupling is at the south pole and we wish to return back to the north pole along a different trajectory characterized $\varphi=\pi/4$. We thus perform a basis rotation from $(-\gamma_1,\gamma_2)$ to a new basis $(-\gamma_1-\gamma_2,-\gamma_1+\gamma_2)/\sqrt{2}$ which corresponds to the zero modes given in Eq.\ (9) for $\theta=\pi$ and $\varphi=\pi/4$. 
Now we can let $\theta$ vary from $\pi$ to $0$, so the system returns back to the north pole. The zero modes are
\begin{align}
\cos \theta \frac{\gamma_1+\gamma_2}{\sqrt{2}}-\sin \theta \Gamma_1,\quad \frac{-\gamma_1+\gamma_2}{\sqrt{2}}.
\end{align}
As $\theta$ varies from $\pi$ to $0$, we thus have
\bee
\gamma_1+\gamma_2\rightarrow -\gamma_1-\gamma_2, \quad
\gamma_1-\gamma_2\rightarrow \gamma_1-\gamma_2
\ee
Performing finally another rotation in $\varphi$ by $-\pi/4$ to return to the original basis we obtain for the overall evolution of $\gamma_1$ and $\gamma_2$
\begin{align}
\gamma_1 &\rightarrow -\gamma_1 \rightarrow  \gamma_2, \\
\gamma_2 &\rightarrow \gamma_2 \rightarrow  -\gamma_1,
\end{align}
showing that the two Majoranas have indeed exchanged in this process.

To understand how this exchange protocol can be implemented using the device depicted in Fig.\ 3e it is easiest to consider the process in reverse. As we further elaborate in Appendix C below it is easy to see that twisting the superconducting phase  $\phi_1$ in the upper half of the wire by $2\pi$ corresponds to the trajectory on the unit sphere going from the north to the south pole with $\varphi=\pi/4$. The underlying physics is captured by Eq.\ (\ref{b1}) below. Now the problem is that although the physical Kitaev Hamiltonian is mapped back onto itself under such a $2\pi$ phase twist, the effective Majorana Hamiltonian $H(s)$ is mapped to $-H(s)$. This is because the definition of the Majorana operators involves $e^{i\phi/2}$ and is therefore not single valued in $\phi$. One can deal with this issue by a redefinition of the Majorana operators in the upper segment of the wire, as discussed e.g.\ in Ref.\ \onlinecite{halperin1}. One can, alternately, imagine undoing the phase twist performed on $\gamma_1$ by twisting the phase of a short segment of the wire very close to its top end by $-2\pi$. This has no effect on $\gamma_2$ and can be pictured as going back to the north pole along the $\varphi=0$ line. Since $\gamma_2$ is not involved in this last (imagined) step it has no effect on braiding and simply implements the transformation back to the original basis. This completes the path indicated in Fig.\ 2b (taken in reverse). 

Comparing the two braiding processes in Fig.\ 2, we find that they have one feature in common. When the paths are projected onto the unit sphere, the covered areas  are $\pi/2$. In the following, we will prove that for an arbitrary closed coupling path, $\gamma_1$ and $\gamma_2$ exchange if the path begins at one of the poles and if the covered area is $\pi/2$.

Following the method employed in Ref.\ \onlinecite{Coupling_braiding}, we compute Berry's phase of the ground states after a coupling cycle. Accumulation of the Berry phase can be regarded as the result of the braiding operation\cite{PhysRevLett.86.268}. Let us rewrite the coupling Hamiltonian in Eq.\ (6) in another economical way 
\bee
H=i(X \gamma_1 + Y \gamma_2 + Z \Gamma_1)\Gamma_2, \label{Hcoupling}
\ee
where $X=E\sin \theta \cos \varphi$, $Y=E\sin \theta \sin \varphi$, and $Z=E\cos \theta$. When the coupling is off, the four Majorana fermions possess zero energy. The ground state has four-fold degeneracy and can be  represented by $\ket{0},\ c^\dag\ket{0},\ d^\dag \ket{0},\ d^\dag c^\dag \ket{0}$, where each fermionic operator is formed by two Majorana operators $c=(\gamma_1-i\gamma_2)/2$ and $d=(\Gamma_1-i\Gamma_2)/2$. The coupling Hamiltonian can be rewritten in this fermionic basis 
\bee
H=
E\bma
Z & 0 & 0 & -X-iY \\
0 & Z & -X+iY & 0 \\
0 & -X-iY & -Z & 0 \\
-X+iY & 0 & 0 & -Z
\ema.
\ee
Due to the conservation of fermionic parity, two blocks of the Hamiltonian with different parities can be discussed separately
\begin{equation}
H_{\rm{even}}=H_{\rm{odd}}^*=
\bma
Z & -X-iY \\
-X+iY & -Z
\ema.
\end{equation}
Turning on the coupling changes the ground state degeneracy from four-fold to two-fold.  The two ground states with energy $-E$ in the even and odd parity sectors are given by
\begin{align}
\ket{\rm{e}}&=
\frac{1}{\sqrt{2E(E-Z)}}
\bma
-E+Z \\
-X+Yi
\ema,\quad 
\\
\ket{\rm{o}}&=
\frac{1}{\sqrt{2E(E-Z)}}
\bma
-E+Z \\
-X-Yi
\ema.
\end{align}
Now we introduce differential forms to compute the Berry phases. The Berry connections ($\bra{\Psi}d\ket{\Psi|}$) in even and odd parity sectors are simply written as differential one-forms\cite{Nakahara:2003ve}
\bee
A_{\rm{even}}=-A_{\rm{odd}}=-\frac{i(XdY-YdX)}{2E(E-Z)},
\ee
and the Berry curvatures ($d\bra{\Psi}d\ket{\Psi|}$), which are differential two-forms, are given by 
\begin{align}
dA_{\rm{even}}&=\frac{i}{2E^3}(ZdX\wedge dY+XdY\wedge dZ+ Y dZ\wedge dX)\nonumber\\
&=\frac{i}{2}\sin \theta d\theta  \wedge d \varphi.
\end{align}
We note that $E$ is not constant so $dE^2=2EdE=2XdX+2YdY+2ZdZ$. After performing a closed loop operation, the original ground states gain extra Berry phases
\begin{align}
\ket{e'}&=\exp\big ( \oint_\mathcal{C}A_{\rm{even}} \big )\ket{e}=\exp\big ( \int dA_{\rm{even}} \big )\ket{e},  \nonumber \\
\ket{o'}&=\exp\big ( \oint_\mathcal{C}A_{\rm{odd}} \big )\ket{o}=\exp\big ( \int dA_{\rm{odd}} \big )\ket{o}. \label{Berry Ground}
\end{align}
On the one hand, the line integrals become surface integrals by Stokes' theorem so $2i\int dA_{\rm{even}}=-2i\int dA_{\rm{odd}}=\int \sin \theta d\theta  \wedge  d\varphi$ is the area covered by the coupling path on the unit sphere.  On the other hand, at the beginning of the process $\theta=0$ so the initial ground states are given by 
\bee
\ket{\rm{e}}=\ket{0} ,\quad \ket{\rm{o}}=c^\dagger \ket{0}
\ee
Ref.\ \onlinecite{PhysRevLett.86.268} shows that when $\gamma_1$ and $\gamma_2$ braiding occurs, 
\bee
\ket{\rm{e}'}=e^{i\pi/4}\ket{\rm{e}},\quad \ket{\rm{o}'}=e^{-i\pi/4}\ket{\rm{o}}.
\ee
Using the relation between the final ground states $\ket{o'}=(\gamma_1'+i\gamma_2')\ket{e'}$, we have $\gamma_1'=\gamma_2$ and $\gamma_2'=-\gamma_1$. Therefore, to achieve braiding between $\gamma_1$ and $\gamma_2$  in the coupling process, the area $\int \sin \theta d\theta \wedge   d\varphi=\pi/2$ is required by comparing the Berry phases in Eq.(\ref{Berry Ground}).

\section{Majorana Josephson junction}
When Majorana modes are present in a Josephson junction, it is known that the current phase relation has an anomalous $4 \pi$ periodicity.\cite{PhysRevB.79.161408} We consider a similar junction device as illustrated by Fig.\ 3e. However, the $4\pi$ periodicity is absent in this device due to the proximity of other Majorana zero modes at the two ends of the wire. In this situation a $2\pi$ phase twist in the bottom half of the wire is equivalent to a braiding operation of Majorana modes $\gamma_1$ and $\gamma_2$ located at the ends. As we demonstrate below this occurs when  $\gamma_1$ and $\gamma_2$ couple to the same linear combination of the Majorana modes located at the junction. 

%Separate to two parts
%1. Majorana phase change after $2\pi$ rotation 
%2. Coupling to effective braiding

The interplay of four Majorana fermions achieves the braiding operation. In the following we denote the Majorana fermion located on the upper (lower) part of the junction by $\alpha$ ($\beta$). To demonstrate the braiding operation, let $\phi_1=0$ remain fixed as $\phi_2=\phi$ varies from $0$ to $2\pi$. The coupling between the two Majorana modes is given by $iE\cos(\phi/2)\Gamma_1\Gamma_2$. As $\phi=0,\ 2\pi$, the two junction Majorana coupling suppresses the coupling effect between the edges ($\gamma_1,\gamma_2$) and junction ($\Gamma_2$). As $\phi=\pi$, the edge and junction coupling dominates in the absence of the junction coupling.   The low-energy effective Hamiltonian can be written as
\bee 
\label{b1}
H_J^{2\pi}(\phi)=i E \big( \epsilon\sin (\phi/2) \frac{\gamma_1+\gamma_2}{\sqrt{2}}+\cos (\phi/2)\Gamma_1)\Gamma_2,
\ee
where $\Gamma_1=-\frac{\alpha-\beta}{\sqrt{2}}$, $\Gamma_2=\frac{\alpha+\beta}{\sqrt{2}}$.  $E(\phi)$ and $\epsilon$ are positive constants. Tuning of a single parameter (such as the wire chemical potential $\mu$) is required to achieve symmetric coupling to a single linear combination of the junction Majoranas. 

When the phase $\phi$ varies, the wavefunctions of $\beta$ and $\gamma_2$ change according to
\bee
\beta(\phi)=e^{i\phi/2}c^\dag_{2j}+e^{-i\phi/2}c_{2j},\quad 
\gamma_2(\phi)=e^{i\phi/2}c^\dag_{2b}+e^{-i\phi/2}c_{2b},
\ee
where $c_{2j}$ and $c_{2b}$ are $\phi$ independent fermionic operators representing $\beta$ and $\gamma_2$ respectively. After the $2\pi$ twist of the bottom half of the wire, $\beta(2\pi)=-\beta(0)$ and $\gamma_2(2\pi)=-\gamma_2(0)$. On the other hand, the evolution of the two zero energy modes of $H_J^{2\pi}(\phi)$ can be written as a function of $\phi$
\bee
\frac{\gamma_1-\gamma_2}{\sqrt{2}},\quad \cos (\phi/2) \frac{\gamma_1+\gamma_2}{\sqrt{2}} - \epsilon\sin (\phi/2)\Gamma_1 \label{evolution J}
\ee
The second Majorana operator is unnormalized for simplicity.  Following the same line or argument as in Sec.\ \ref{generalization} above, after $2\pi$ rotation, we find that  
\begin{align}
\gamma_1 &\rightarrow -\gamma_2(2\pi)=\gamma_2(0)\\
\gamma_2 &\rightarrow -\gamma_1
\end{align}
We thus observe that braiding of $\gamma_1$ and $\gamma_2$ can be achieved by twisting  the phase in one half of the wire by $2\pi$. In this process the system comes back to the original ground state (up to an important overall phase). This is unlike the Josephson effect with $4\pi$ periodicity. In the following, we compare these two different processes. 

\begin{figure*}[t]
 \begin{center}
   \subfigure[$\mu=0.185t$]{\label{fig:edge-a}\includegraphics[width=55mm]{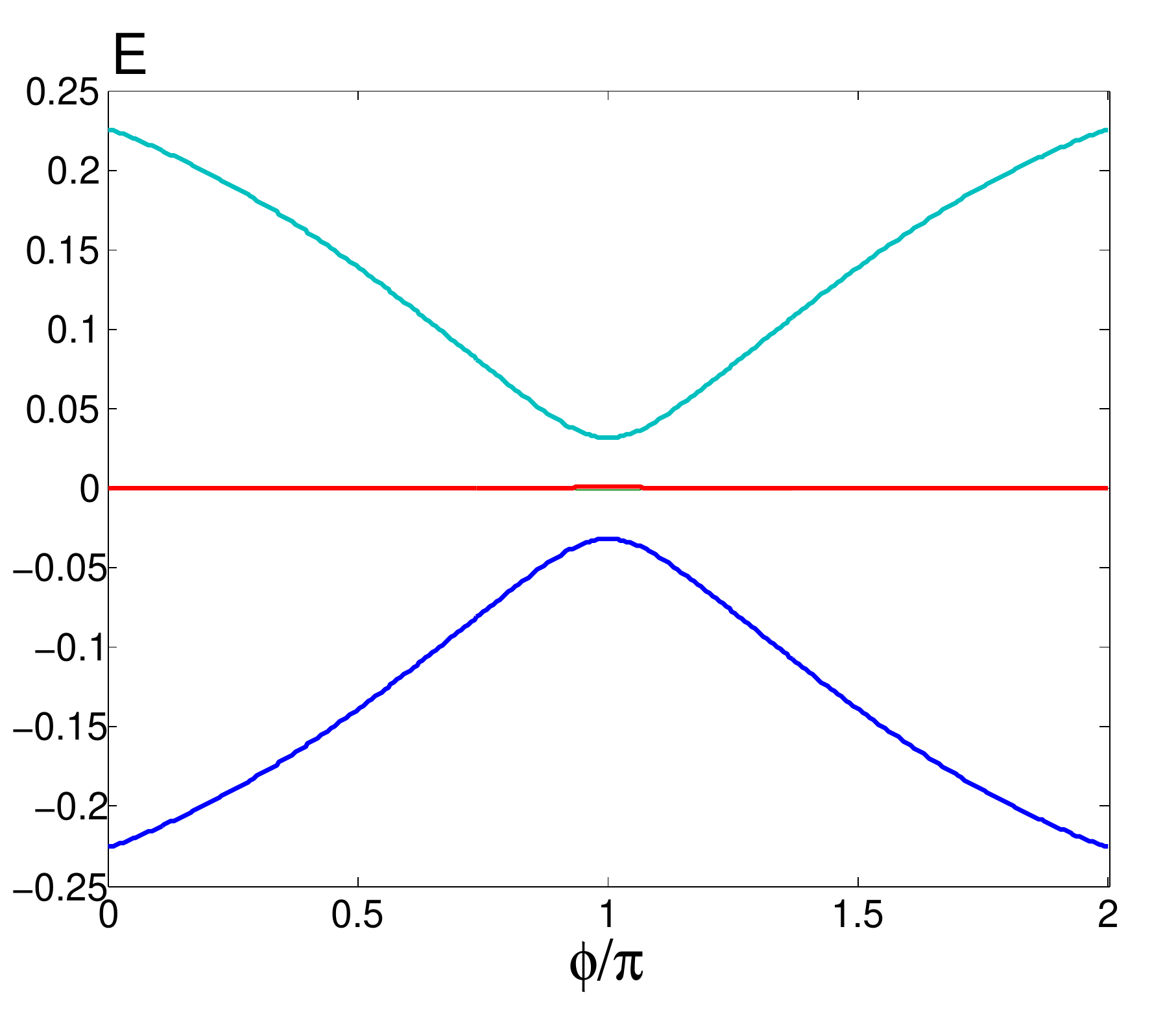}}
   \subfigure[$\mu=0.14t$]{\label{fig:edge-b}\includegraphics[width=55mm]{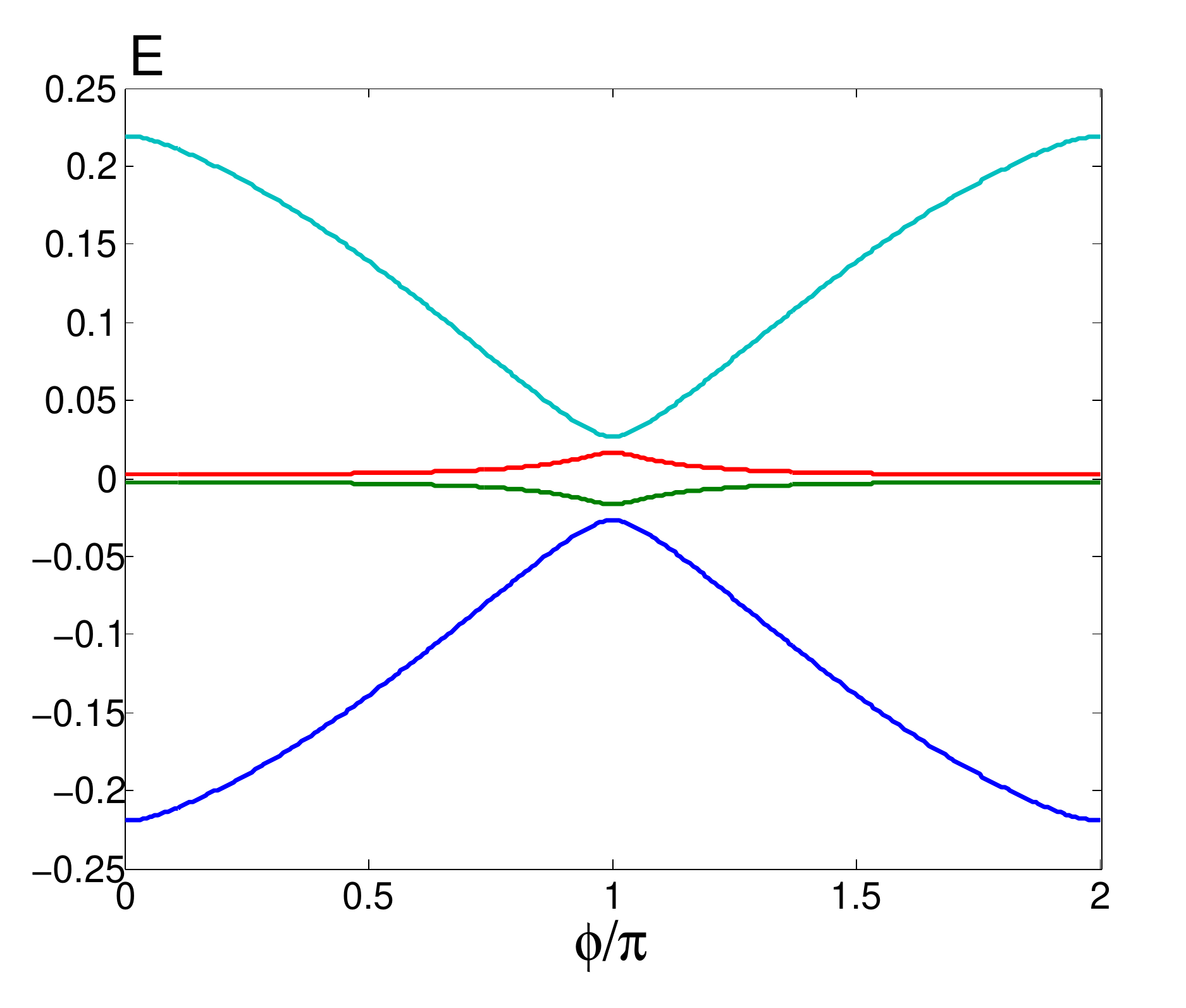}}
      \subfigure[The couplings of the Majorana modes as $\mu=0.185t$]{\label{fig:edge-c}\includegraphics[width=55mm]{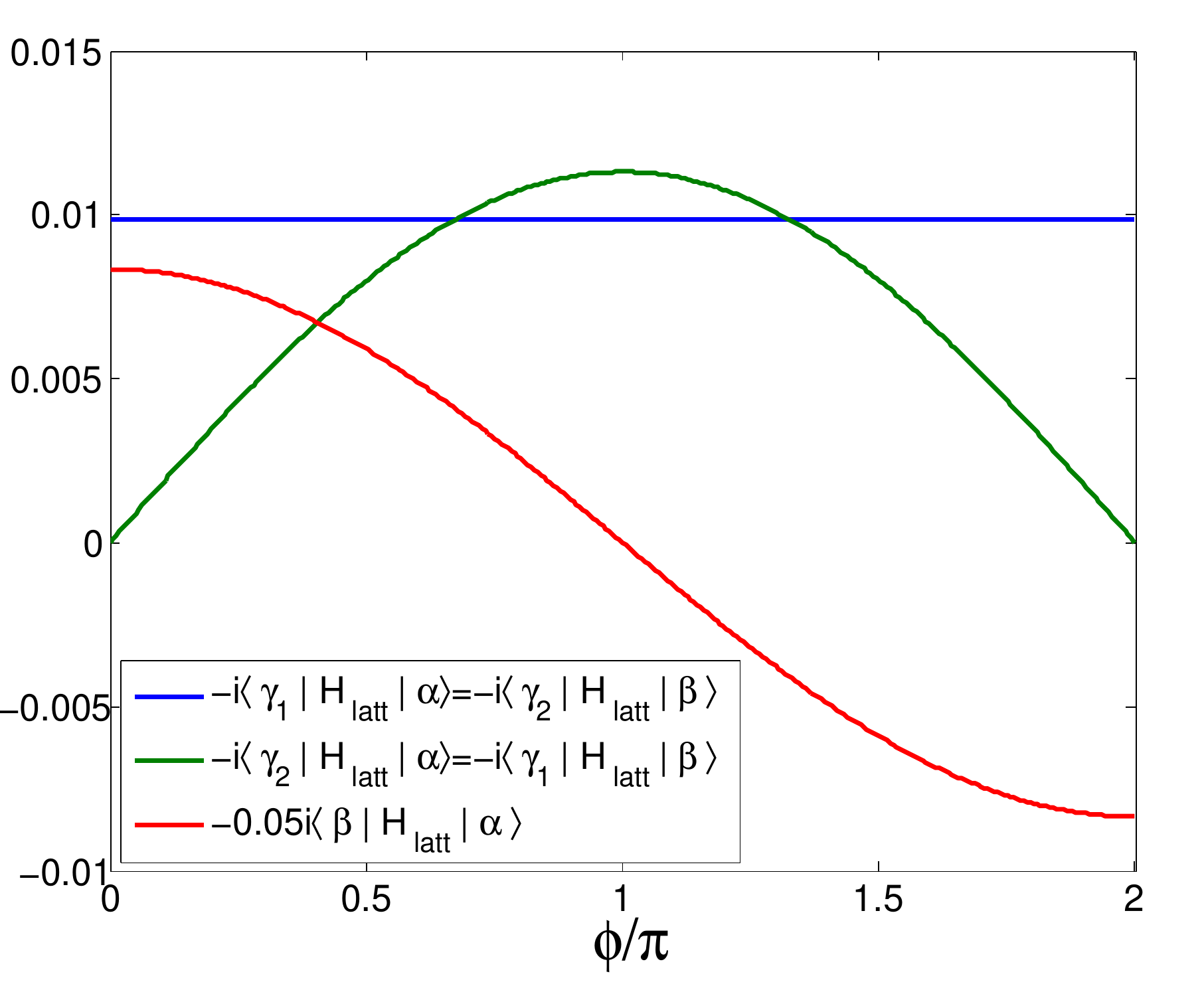}} \\
 \end{center}
 \caption{The lowest energy eigenvalues of $\cal H_{\rm latt}$ in Eq.\ (12) with $\Delta=0.1t$ and $N=50$.}
 \label{delta}
\end{figure*}

For the $4\pi$ Josephson junction, we can imagine that the two ends of the chain are at infinity so $\gamma_1$ and $\gamma_2$ are completely decoupled and remain exact zero modes throughout. In this regard, the Hamiltonian is given by only the coupling of two Majorana fermions at the junction 
\bee
H_J^{4\pi}(\phi)=iE\cos(\phi/2)\beta \alpha
\ee
Both of these two processes start from the same ground states ($\phi=0$) obeying 
\bee
(\beta+i\alpha)\ket{G_{2\pi,4\pi}}=0
\ee 
with the energy $-E$. We follow the evolution of the two gapped Majorana operators. In the $4\pi$ junction regime, the two gapped Majorana modes $(\alpha,\ \beta)$ stay in the same form. In the braiding regime the evolution of the Majorana modes is given by 
\bee
\epsilon\sin (\phi/2) \frac{\gamma_1+\gamma_2}{\sqrt{2}}+\cos (\phi/2)\Gamma_1,\ \Gamma_2,
\ee
The final modes at $\phi=2\pi$, which are $\Gamma_1\rightarrow -\Gamma_1$ and $\Gamma_2\rightarrow \Gamma_2$, imply $\alpha\rightarrow \beta$ and $\beta\rightarrow\alpha$. 
On the other hand, due to the $2\pi$ phase rotation, $\beta\rightarrow -\beta$ in both cases. The ground states in the braiding protocol and the $4\pi$ Josephson junction satisfy two different equations
\bee
-i(\beta+i\alpha)\ket{G'_{2\pi}}=0,\quad -(\beta-i\alpha)\ket{G'_{4\pi}}=0
\ee
At the same time, the Hamiltonian evolves to the original Hamiltonian 
\bee
H_J^{2\pi,4\pi}=-iE\beta\alpha(2\pi)=iE\beta\alpha
\ee
The ground state of the braiding Hamiltonian stays the same with the energy $-E$ but the ground state of the $4\pi$ Josephson junction evolves to the excited state with the energy $E$. Thus, the presence or absence of the coupling to Majorana fermions at the ends of the wire brings two completely different outcomes --- braiding and $4\pi$ periodicity, respectively. 

We now support the above analysis by a detailed calculation using the lattice model $\cal H_{\rm latt}$ of the TSC chain defined in Eq.\ (12).
We show that the Hamiltonian $H^{2\pi}_J$ in Eq.\ (\ref{b1}) indeed describes the low-energy degrees of freedom in the one-dimensional TSC modulo a small correction that we argue is unimportant for the outcome of the braiding operation. To this end we numerically solve the Hamiltonian $\cal H_{\rm latt}$ with the order parameter distribution given by
\bee
\Delta_{j,j+1}=
\Biggl\{
\begin{matrix}
\Delta,\quad \quad  \ j<M \\
0,\quad \quad\ j=M  \\
\Delta e^{i\phi},\ \ j>M \\
\end{matrix},
\ee
where $M=N/2$ (taking $N$ even)  and the other parameters are the same as in Fig.\ 3. As illustrated by Fig.\ \ref{delta} the energy eigenvalues of $\cal H_{\rm latt}$ as a function of $\phi$ show gapped and gapless properties in two distinct situations. In Fig.\ \ref{delta}a, two of the Majorana modes remain at zero energy and the absence of the energy level crossing indicates the absence of the ground state switching. This  low-energy spectrum is consistent with the form expected for $H_J^{2\pi}(\phi)$ and we thus tentatively conclude that braiding occurs for these parameters. The energy levels in Fig.\ \ref{delta}b show a significant splitting between the zero modes. This indicates that additional terms not included in $H_J^{2\pi}(\phi)$ are present in the low-energy theory. In this situation we do not expect braiding to occur; rather, if the phase twist is implemented relatively fast we expect a transition to the excited state and the resulting $4\pi$-periodic behavior.

To further confirm that the effective low-energy theory of $\cal H_{\rm latt}$ is described by $H_J^{2\pi}$, we now compute the effective couplings between the four Majorana fermions $(\gamma_1,\ \gamma_2,\ \alpha,\ \beta)$.
First, imagine a cut in the middle of the chain, that is, we let $t=\Delta=0$ on the link between sites $M$ and $M+1$ in $\cal H_{\rm latt}$. Next, by using the values of the parameters in Fig.\ \ref{delta}a and then solving the eigenvalue problem, we obtain the wavefunctions of the four zero energy modes at the ends of the two separated chains. We note that $\gamma_2$ and $\beta$, which are $\phi$-dependent, must be computed each time while $\phi$ varies. Second,  restoring non-zero $t$ on the middle link we sandwich $\cal H_{\rm latt}$ between these four Majorana mode wavefunctions and obtain the relevant  couplings. These are shown in Fig.\ \ref{delta}c. 

On the basis of these results, the effective low-energy Hamiltonian can be written as a function of $\phi$
\begin{align}\label{heff}
H_{\rm eff}(\phi)=i\frac{E}{2} &  \big (\epsilon' \gamma_1\alpha +  \epsilon'\gamma_2 \beta 
+ \epsilon\sin(\phi/2)\gamma_1\beta \nonumber\\
&+ \epsilon \sin(\phi/2) \gamma_2 \alpha+ 2 \cos(\phi/2)\beta\alpha \big )
\end{align}
We note that $\epsilon' \sim \epsilon \ll 1$ because the coupling between $\beta$ and $\alpha$ is usually much larger than couplings between distant Majoranas. The effective Hamiltonian can be rewritten in an economical way 
\begin{align}
H_{\rm eff}(\phi)=H^{2\pi}_J(\phi)+i\frac{E}{2}(\epsilon'-\epsilon\sin (\phi/2))( \gamma_1\alpha+\gamma_2\beta) \label{correction}
\end{align}
Thus, the low-energy sector of the Kitaev lattice Hamiltonian with a phase twist is  described by $H_J^{2\pi}$ with an extra term. When $\epsilon'\neq\epsilon$ the extra term produces a splitting between the zero modes $\delta E\simeq E|\epsilon'-\epsilon|$ for $\phi=\pi$. We can thus identify the extra terms in $H_{\rm eff}(\phi)$ as being responsible for the behavior indicated in Fig.\ \ref{delta}b. As in the case of the Majorana shuttle,  $\epsilon'$ and $\epsilon$  depend on the system parameters. Fig.\ \ref{delta}a shows that by tuning a single parameter, e.g.\ the chemical potential $\mu$, we can achieve the situation in which $\epsilon'=\epsilon$ and Majorana zero modes remain robust during the process.

\bibliographystyle{apsrev}